# Genetic Algorithm-Guided Deep Learning of Grain Boundary Diagrams: Addressing the Challenge of Five Degrees of Freedom

Chongze Hu, Yunxing Zuo, Chi Chen, Shyue Ping Ong, Jian Luo[*]

*Department of Nanoengineering; Program of Materials Science and Engineering, University of California San Diego, La Jolla, California 92093, USA*

**Abstract**

Grain boundaries (GBs) often control the processing and properties of polycrystalline materials. Here, a potentially transformative research is represented by constructing GB property diagrams as functions of temperature and bulk composition, also called "complexion diagrams," as a general materials science tool on par with phase diagrams. However, a GB has five macroscopic (crystallographic) degrees of freedom (DOFs). It is essentially a "mission impossible" to construct property diagrams for GBs as a function of five DOFs by either experiments or modeling. Herein, we combine isobaric semi-grand-canonical ensemble hybrid Monte Carlo and molecular dynamics (hybrid MC/MD) simulations with a genetic algorithm (GA) and deep neural network (DNN) models to tackle this grand challenge. The DNN prediction is ~$10^8$ faster than atomistic simulations, thereby enabling the construction of the property diagrams for millions of distinctly different GBs of five DOFs. Notably, excellent prediction accuracies have been achieved for not only symmetric-tilt and twist GBs, but also asymmetric-tilt and mixed tilt-twist GBs; the latter are more complex and much less understood, but they are ubiquitous and often limit the performance properties of real polycrystals as the weak links. The data-driven prediction of GB properties as function of temperature, bulk composition, and five crystallographic DOFs (*i.e.*, in a 7D space) opens a new paradigm.

---

[*] Corresponding author. E-mail: jluo@alum.mit.edu

**Introduction**

Most engineered and natural materials are polycrystalline, where grain boundaries (GBs) can often control a variety of properties. Like a three-dimensional (3D) bulk phase, a GB can adopt an equilibrium state for a given set of thermodynamic variables such as temperature, pressure, and chemical potential [1]. Such a GB equilibrium state can be treated as an interfacial phase that is thermodynamically 2D, which is also named as "complexion" to differentiate it from a thin layer of 3D phase precipitated at the GB [1-3]. Impurities or solutes adsorption (*a.k.a.* segregation) at GBs, which can occur along with GB structural transformations [1], can drastically change microstructural evolution [3] or cause catastrophic embrittlement [4, 5]. More generally, GB adsorption and structures can affect a broad range of kinetic, mechanical, electronic/ionic, magnetic, thermal, and other properties [1, 6]. Thus, understanding the GB composition-structure-property relation is of both fundamental and practical interest.

Phase diagrams, maps of thermodynamically stable (3D bulk) phases as functions of temperature and composition, are one of the most important materials science tools. Since the properties of GBs can be as important as those of the bulk phases in polycrystalline materials, it can be broadly useful to map out the GB states and properties as functions of temperature and bulk composition, thereby constructing the GB counterparts to bulk phase diagrams [7-13]. Some computed GB diagrams include first-order transformation lines [7-9, 11], while GB transformations are continuous in many other materials. In both cases, constructing GB property diagrams as functions of temperature and bulk composition can be highly useful.

However, a GB has five macroscopic (crystallographic) degrees of freedom (DOFs) [14, 15], in addition to several thermodynamic DOFs, including temperature and the chemical potential(s) set by the bulk composition. Thus, it becomes virtually a "mission impossible" to construct bulk composition- and temperature-dependent property diagrams for GBs as functions of five crystallographic DOFs by either experiments or modeling. Advanced electron microscopy [16-18] and atom probe tomography [19] have been used to characterize GB structures at the atomic level, but the sample preparation and experimental procedures are time-consuming even for studying a small number of GBs at one given thermodynamic condition. Prior works of computing GB diagrams as functions of temperature and bulk composition mostly used highly simplified phenomenological or lattice models [7, 10-12]. A WC-Co interfacial diagram was constructed via a DFT-based cluster-expansion method that only surveyed a limited number of given interfacial structures [20]. Only a couple of recent studies computed GB diagrams as functions of the bulk composition (that sets/represents the chemical potential) and temperature via atomistic simulations, but only for one special tilt or twist GB [8, 9]. Recently, various advanced computational methods, such as the evolutionary algorithm [21], genetic algorithm (GA) [8, 22], Bayesian optimization [23], and machine learning (ML) [24-26], combined with high-throughput calculations [27], data mining [28], and virtual screening [29], have been used to model GBs; however, most of these studies focused on either symmetric-tilt or twist GBs. More general and asymmetric GBs [16], which are ubiquitous in polycrystals and often the weak links chemically and mechanically, are hitherto scarcely investigated by virtually any modeling method. Understanding the GB properties as functions of seven DOFs (five crystallographic DOFs plus bulk composition and temperature) remains a grand challenge, which motivated this study.

Herein, we demonstrate a GA-guided deep learning approach to predict GB properties as a function of seven DOFs (or construct bulk composition- and temperature-dependent GB diagrams as function of five DOFs) by combining isobaric semi-grand-canonical ensemble hybrid Monte Carlo and molecular dynamics (hybrid MC/MD) simulations, GA-based variable selection, and deep neural networks (DNN) prediction. Notably, our final DNN model is capable of predicting properties of not only simpler symmetric-tilt and twist GBs, but also more complex and general GBs.

**Results and Discussions**

**Workflow of GA-guided deep learning of GB diagrams**

The workflow of the GA-guided DNN prediction for GB diagrams is displayed in Fig. 1. The detailed procedure is given in §1 Methods in Supplementary Material (SM). The Cu-Ag system is selected as our model system because of its robust interatomic potential [30].

First, large-scale isobaric semi-grand-canonical (*i.e.*, constant-$N\Delta\mu PT$) ensemble hybrid MC/MD simulations were performed to compute three GB properties (*i.e.*, GB adsorption $\Gamma_{\text{Ag}}$, GB excess disorder $\Gamma_{\text{Disorder}}$, and GB free volume $V_{\text{Free}}$) for 30 symmetric-tilt (ST), 30 twist (TW), 18 asymmetric-tilt (AT), and 22 mixed tilt-twist (MX) GBs as a function of chemical potential difference $\Delta\mu$ (= $\mu_{\text{Ag}} - \mu_{\text{Cu}}$) and temperature $T$ (Fig. 1a). All together, we performed 6,581 individual constant-$N\Delta\mu PT$ atomistic simulations (each with ~16,000-50,000 atoms for ~1 million hybrid MC/MD steps at constant $N\Delta\mu PT$ until reaching convergence). To the best of our knowledge, the simulations here have generated the largest and most systematic dataset of binary GBs to date, which are used for the subsequent data-driven prediction of GB diagrams as a function of five macroscopic DOFs.

Second, the dataset generated by hybrid MC/MD simulations was used to perform GA-based variable selection to identify the most significant GB descriptors for each property and each GB type, and subsequently for all GBs together in a unified DNN model (Figs. 1b-c).

Third, the GA-selected descriptors were used to train, validate, and test DNN models (Fig. 1d). Finally, the DNN models were used to predict GB diagrams as functions of five DOFs plus two thermodynamic DOFs (Fig. 1e).

**Benchmark of atomistic simulations**

To ensure the quality of our data generated by large-scale atomistic simulations, we performed comprehensive benchmark tests of both *NPT* MD simulations of pure Cu and $N\Delta\mu PT$ hybrid MC/MD simulations of Ag-doped Cu. The embedded-atom method (EAM) potential developed by William et al. [30] was adopted. Overall, our simulations are consistent with prior experimental [19, 31] and modeling [21, 32, 33] results. Selected results are discussed here with more details in §2.1 and §2.2 in SM. First, the GB energy ($E_{GB}$) and free volume ($V_{Free}$) obtained from the *NPT* simulations agree with the first-principles DFT calculations; in particular, the $V_{Free}$ values obtained from the *NPT* simulations agree better with DFT results than those from the *NVT* simulations [32]. Second, the $N\Delta\mu PT$ simulated bulk phase diagram agrees well with the experimentally-measured phase diagram [34]. Third, temperature-induced GB structural transformations from normal to split and filled kites are evident, being qualitatively consistent with prior *NVT* simulations [33]; moreover, *NPT*-simulated GB structures *vs.* misorientation angles are also consistent with prior *NVT* simulations with open surfaces [21, 32]. Fourth, both *NPT* and $N\Delta\mu PT$ simulated GB structures and properties are robust with respect to perturbations of adding vacancies and interstitials, thereby demonstrating that the GB free volume (local density) can be effectively adjusted in our constant-pressure simulations to achieve equilibria. Fifth, our $N\Delta\mu PT$ simulations of GB adsorption *vs.* bulk $X_{Ag}$ agree with prior MC simulations for a Σ5 GB [19, 33] within our simulation errors. See more detailed discussions in §2.1, §2.2, Figs. S1-S6, and Table S1 in SM.

It is worth noting that the $N\Delta\mu PT$ hybrid MC/MD simulations show continuous transformations of GB adsorption, disordering, and free volume with temperature or bulk composition, which are discussed and critically compared with prior studies in SM §2.3. Notably, our $N\Delta\mu PT$ hybrid MC/MD simulations successfully reproduced the segregation-induced GB nano-faceting in an asymmetric GB observed in a prior experiment [31] (§2.2.4 and Fig. S8 in SM). The rationales for selecting the current simulation method and three GB properties are discussed in SM §2.4.

After benchmarking and validating our methods, we conducted 6,581 MC/MD simulations to collect a large dataset. Subsequently, we performed GA to identify the important GB descriptors and developed DNN models using this dataset.

**Genetic algorithm (GA) to select significant GB descriptors**

GBs can be characterized by five macroscopic DOFs based on coincident-site-lattice (CSL) misorientation scheme or interface-plane scheme [14, 15, 35]. We primarily adopt the latter, where GBs are classified into four groups: ST, TW, AT, and MX GBs (Fig. 2a) [16]. Moreover, we identified and summarized 38 commonly used GB descriptors in Table S2 in SM, including both structural and geometrical parameters. See the definitions and discussion in Method §1.1 in SM. Note that redundant descriptors are used as input for DNN models for better predictions.

Subsequently, we used GA-based variable selection (Fig. 1b) to identify the significant GB descriptors that control each GB property of each group, as well as all GBs together (Method §1.6 in SM). The GA inspired by Darwinian evolution is a metaheuristic optimization method that can be used to select optimal parameters [36, 37]. One GA process can include variable evaluation, selection, crossover, and mutation to identify the significant descriptors that control the targeted property (Fig. 1b). For ST, TW, and AT GBs, we only consider the first 32 descriptors because these GBs only require either a tilt or a twist rotation to transform Grain 1 to Grain 2. For MX and all-included GBs, all 38 descriptors, including the last six descriptors related to the overall (combined tilt-twist) rotation axis and plane, are considered. We plot the GA scores of significances for these GB descriptors in Fig. 2b and use red pentagrams to label the significant GB descriptors.

The GA-selected significant GB descriptors for ST, TW, AT, TW, and all-included GBs are discussed in detail in SM §2.5. Selected interesting findings are outlined here. First, the GA identified Σ, the inverse of degree of coincidence, as a significant descriptor for ST, TW, and AT GBs. An important result is that Σ was not selected as a significant descriptor for the most general MX GBs. This result challenges the class wisdom [14] by suggesting that Σ is not an important parameter to characterize the properties of general GBs; however, this maverick finding is supported by a series of recent studies by Rohrer, Randle, and their co-workers [38-41]. Second, the GA found the orientations of the terminal GB planes to be important in general, which is consistent with recent theoretical [14, 42, 43] and experimental [14, 16, 18, 44] studies. Third, the GA found that denser (lower Miller index) terminating grain plane is more important to control adsorption and structures of asymmetric AT and MX GBs, as well as all-included GBs. Notably, this finding is again consistent with recent advanced microscopy observations that lower-index grain surface dictates faceting and segregation behaviors [16, 44].

Interestingly, the GA also reproduced some common knowledge and conventions. For ST GBs, the GA-identified significant descriptors are all in the CSL-misorientation notation $\varphi_{tlt}$ [$u_{tlt}$ $v_{tlt}$ $w_{tlt}$] ($h$ $k$ $l$) commonly-used to denote ST GBs, where $\varphi_{tlt}$ and [$u_{tlt}$ $v_{tlt}$ $w_{tlt}$] are the tilt angle and axis, and ($h$ $k$ $l$) is the boundary plane [14]. Twist GBs only have five GA-selected significant descriptors, which are all in their characteristic structural relation of $\Sigma = \gamma(\Phi_{twst})\delta r^2$, where $\Phi_{twst}$ is the twist angle (noting that $\Phi_{twst} = \theta_{mis}$ for TW GBs), $\gamma$ is the inverse planar coincident-site density [14], and $\delta = 1$ or $0.5$ to ensure Σ to be odd.

For the most comprehensive case of all-included GBs, a large number of fifteen significant GB descriptors were chosen by the GA. Several interesting findings are discussed as follows. First, the GA selection indicates that $\Sigma^{-1}$ is a significant GB descriptor, and it is better than Σ itself that is not a significant descriptor. Second, the orientations of terminating grain planes are still important. Third, misorientation angle $\theta_{mis}$ was not selected by the GA, but decomposed tilt angle $\varphi_{tlt}$ was chosen. Fourth, the interplanar distance $d_1$ (for the lower Miller index grain surface) was selected (and is more important than $d_2$), which again suggests that the lower-index plane is important overall. Fifth, the overall (combined tilt and twist) rotation axis and plane are also important for all-included GBs. This overall rotation is reduced to the tilt or twist rotation for simpler ST, TW, and AT GBs so they are not present there. It is interesting to note that the inclusion of the overall rotation axis/plane has significantly improved the prediction accuracies for all-included GBs.

We summarized the significant descriptors for each GB property for different GB types in Table S3 in SM. Moreover, we plotted the GA scores of the first 32 descriptors *vs*. GB type in Fig. S10 in SM. The significant GB descriptors for each of four types of GBs, as well as for all GBs combined together, are discussed in detail in SM §2.5.

**Deep neural networks (DNN) model selection and performance**

Using GA-selected descriptors, two-layer single-task DNN models were developed for four groups of (ST, TW, AT, and MX) GBs; a unified DNN model was also developed for all four types together (referred to as "all-included GBs" and labeled as "All" for brevity). The detailed DNN architecture is shown in Fig. 1d. To compare the performance of these five DNN models, we plot the root-mean-square errors (RMSEs) between the MC/MD simulations and DNN predictions for three GB properties ($\Gamma_{Ag}$, $\Gamma_{Disorder}$, and $V_{Free}$) in Fig. 3.

Notably, the three GB properties predicted by the all-included DNN model are sufficiently accurate (with comparable RMSEs with four individual DNN models). Specifically, the unified all-included DNN model in fact works better than the individual DNN models for predicting ST and AT GBs (except for $\Gamma_{Disorder}$), while it performs slightly lower for TW and MX GBs (Fig. 3). Nevertheless, the differences in RMSEs among the four individual DNN models and the unified all-included DNN model are relatively small (Fig. 3 and Fig. S11). Thus, while the GA selection of significant descriptors for each GB type has provided useful physical insights, our subsequent analysis and discussion are focused on the unified all-included DNN model for the sake of its simplicity and generality.

For the unified all-included DNN model, the parity plots between DNN-predicted and MC/MD-computed values of $\Gamma_{Ag}$, $\Gamma_{Disorder}$, and $V_{Free}$, respectively, are shown in Fig. 4a, Fig. 4b, and Fig. 4c, respectively. The linear relations between training, validation, and test datasets demonstrate the robustness of this DNN model. We note that there are relatively large deviations between the DNN predictions and MC/MD simulations at high values, which can be ascribed to high levels of thermal noises in the MC/MD simulations at high temperatures and/or near solidus lines (SM §1.11).

Moreover, we compare the image similarity between DNN-predicted and MC/MD-simulated GB diagrams by calculating the structural similarity index (SSIM; 0 = different and 1 = same; see the definition in Method §1.10). The SSIM histogram plots of three property diagrams ($\Gamma_{Ag}$, $\Gamma_{Disorder}$, and $V_{Free}$) for four groups of ST, TW, AT, and MX GBs, as well as all-included GBs, are shown in Figs. 4d, 4e, and 4f. Relatively high SSIM values of ~0.84-0.93 have been achieved for three GB property diagrams for all-included DNN model, as well as for each of four individual groups of GBs (Figs. 4d to 4f).

As an example, a comparison plot of MC/MD-simulated *vs*. DNN-predicted diagrams for a Σ81 MX GB with boundary planes $(1\bar{1}0)//(7\bar{8}7)$ is shown in Figs. 4g to 4i, where high SSIM values of 0.99, 0.98, and 0.94 are obtained for the $\Gamma_{Ag}$, $\Gamma_{Disorder}$, and $V_{Free}$ diagrams, respectively. The high similarities between MC/MD-calculated and DNN-predicted diagrams further verify the accuracy and robustness of this DNN model. It is worth noting that the relatively lower SSIM values for the $\Gamma_{Disorder}$ and $V_{Free}$ diagrams can be ascribed to their more complex quantification procedures that introduce more noises in the DNN training and test sets.

Using this unified all-included DNN model, we further plotted and documented the MC/MD-calculated *vs*. DNN-predicted the GB diagrams of three properties ($\Gamma_{Ag}$, $\Gamma_{Disorder}$, and $V_{Free}$) for 103 different GBs (*i.e.* 618 GB diagrams), including three additional GBs with Σ > 100 discussed subsequently, in Figs. S15-S117 in SM.

Finally, we tested and compared the speed performance of the DNN predictions *vs*. MC/MD simulations. We tabulated simulation time for one MC/MD simulation with one-million steps by using two Intel microprocessors in Table S4 in SM. On the one hand, the needed CPU hours increase linearly from ~200 to 550 with the increasing size of the simulation cell from ~16,000 to 47,000 atoms. On the other hand, the all-included DNN model only takes ~0.001 CPU seconds for one prediction, which is about $10^8$ faster than a typical MC/MD simulation.

**The effective range of the descriptors for the DNN model**

The DNN model adopted here makes predictions under a certain range of GB descriptors. For example, we selected the Σ value to be between 5 and 99. Here, we excluded Σ = 3 GBs because they are too special with distinctly different behaviors [45]; thus, they should be treated separately. Similarly, we did not consider small-angle (<10º) GBs that also behave differently from more general GBs and are better to be treated separately using dislocation-based models [46]. In other words, the current work focuses on more general GBs (and it actually performs better for more general GBs, as we will show subsequently). We have listed the effective ranges of all 38 GB descriptors used in this study in Table S2 in SM.

**Further extensibility and transferability of the DNN model**

We can further extend the upper bound of the $\Sigma$ value to represent even more general GBs, *e.g.*, those GBs with $\Sigma > 500$ [16]. Such a large $\Sigma$ value renders MC/MD simulation infeasible to provide a training set for the DNN model. Thus, we need a method for extrapolation. The GA selection of all-included GBs shows that $\Sigma$ itself is not important, but the descriptor $\Sigma^{-1}$ becomes significant (Fig. S10 in SM). Since $\Sigma^{-1}$ only varies in a small range beyond $\Sigma > 100$, we can use the DNN model to predict more general GBs via extrapolation.

To test this extensibility, we computed the $\Gamma_{Ag}$, $\Gamma_{Disorder}$, and $V_{Free}$ diagrams for a few GBs with $\Sigma > 100$, including $\Sigma113$, $\Sigma171$ and $\Sigma599$ GBs, as well as an asymmetric (110)//(610) GB characterized in a prior experiment [31]. Since the asymmetric (110)//(610) GB is not a CSL GB ($\Sigma \to \infty$), we conducted a MC/MD simulation on a large cell (with a small strain to allow periodic boundary condition; see Fig. S8 in SM). We used two large $\Sigma$ values (99 and 599) in the DNN predictions to extrapolate (noting that $\Sigma > 500$ represents sufficiently general GBs [16]) to $\Sigma \to \infty$.

First, we found that RMSEs of the three GB properties ($\Gamma_{Ag}$, $\Gamma_{Disorder}$, and $V_{Free}$) predicted by DNN for these four GBs ($\Sigma99$, $\Sigma113$, $\Sigma171$, and $\Sigma599$) still follow the linear trends in the parity plots of three diagrams, as shown in Figs. 3a to 3c. Second, the direct comparisons of MC/MD-simulated *vs*. DNN-predicted GB diagrams shown in Figs. S114-117 in SM also suggest similar similarities and acceptable SSIM values between DNN predictions and MC/MD simulations for these four cases of $\Sigma \geq 99$.

Furthermore, we plotted the GB $\Gamma_{Ag}$, $\Gamma_{Disorder}$, and $V_{Free}$ diagrams predicted from the DNN model using $\Sigma = 99$ and $\Sigma = 599$, along with MC/MD-simulated asymmetric (110)//(610) GB ($\Sigma \to \infty$), in Fig. S12 in SM for comparison. Notably, the three GB diagrams predicted by the DNN model using $\Sigma = 99$ and $\Sigma = 599$, respectively, are essentially identical (SSIM = 1.0 for all comparisons; see Fig. S12 in SM). Moreover, these DNN-predicted GB diagrams are similar to the MC/MD-simulated GB diagrams with high SSIM values of 0.96, 0.86, and 0.83, respectively, for the GB $\varGamma_{\text{Ag}}$, $\varGamma_{\text{Disorder}}$, and $V_{\text{Free}}$ diagrams, respectively.

Thus, we conclude that this all-included DNN model can also predict properties for GBs with $\Sigma > 100$ from extrapolation by adopting $\Sigma^{-1}$ as an input parameter.

The prior experiment [31] found Ag segregation induced nano-faceting at this (110)//(610) GB, which has also been successfully reproduced by our MC/MD simulations (Fig. S8, §2.2.4 in SM). This suggests that the DNN prediction is also valid for nano-faceted GBs.

It is worth noting that other significant GB descriptors (as the input of the DNN model) also have effective ranges (given in Table S2 in SM). Especially, some descriptors (*e.g.*, $h_1$ and $k_2$) may go beyond the bound when $\Sigma$ is set to a too large value. Nonetheless, GBs with $\Sigma \sim 100$ (or $\Sigma^{-1} \sim 0.01$) are typically sufficiently good to represent general GBs, *e.g.*, the case shown in Fig. S8 in SM.

To apply the current method to another alloy system, we have to complete a substantial amount of MC/MD simulations to feed the DNN model. Yet, this work still represents a breakthrough because it was considered a "mission impossible" to map our GB properties as function of temperature, bulk composition, and five crystallographic DOFs in a 7-D space prior to the development of this method. In addition, the large dataset generated in this work enables future works to develop new and better phenomenological models with a few parameters. If that is successful, it may enable us to predict useful trends with a much smaller set of simulations.

**Comparison of 300 Pairs of MC/MD-Simulated *vs.* DNN-Predicted GB Diagrams**

We further examine the image similarities of 300 pairs of MC/MD-simulated *vs.* DNN-predicted GB diagrams by analyzing the SSIM distributions as functions of the GB type, symmetry, and selected key GB descriptors.

The SSIM distributions for three GB properties calculated for 100 GBs are plotted in the space of misorientation angle $\theta_{\text{mis}}$ and twist angle $\varPhi_{\text{twst}}$ in Fig. 5a to 5c. It shows that ST and AT GBs ($\varPhi_{\text{twst}} = 0$) have relatively low SSIM values when tilt-rotation angles are small, as indicated by blue arrows in Fig. 5c. This is because small-angle tilt GBs behave differently from the other more general GBs that we used to train the DNN model, and they can be better represented by dislocation-based models.

In addition, some TW GBs ($\theta_{\text{mis}} = \varPhi_{\text{twst}}$), especially (111) twist GBs, also exhibit relatively low SSIM values because they have very small values of $\Gamma_{\text{Ag}}$, $\Gamma_{\text{Disorder}}$, and $V_{\text{Free}}$ (so the relative percentage errors are larger) and they are also more special GBs. Again, such special (high-symmetry and low-energy) GBs can be more effectively modeled by other conventional methods.

Notably, MX GBs ($\theta_{\text{mis}} \neq \Phi_{\text{twst}}$) exhibit high SSIM values (as indicated by purple ellipses in Fig. 5a to 5c) because these GBs are more general GBs.

Moreover, the SSIM distribution for three GB properties calculated for 100 GBs are plotted in the space of interplanar distances $d_1$ and $d_2$ in Fig. 5f to 5f. It clearly shows that some GBs with high symmetry ($d_1 = d_2$) exhibit relatively low SSIM values (indicated by blue arrows), while the more general asymmetric GBs ($d_1 \neq d_2$) are better predicted by the DNN model with high SSIM values.

Here, an interesting and useful finding is that our unified all-included DNN model performs better for more general (particularly asymmetric mixed) GBs, which are more important but much less understood by existing theories and more difficult to model by other conventional methods. The DNN model performs relatively less well for some more special GBs, such as small-angle tilt GBs and high-symmetry (111) twist GBs, which can be better modeled by conventional methods. Note that all the analyses here are consistent with the SSIM histograms shown in Fig. 4. See more discussion in SM §2.6.

In summary, this unified all-included DNN model can predict the temperature- and bulk composition-dependent $\Gamma_{\text{Ag}}$, $\Gamma_{\text{Disorder}}$, and $V_{\text{Free}}$ diagrams for an extremely large number of distinct ST, TW, AT, and MX GBs in the space of five macroscopic DOFs. In other words, this DNN model can map out the $\Gamma_{\text{Ag}}$, $\Gamma_{\text{Disorder}}$, and $V_{\text{Free}}$ values in a 7-D space!

Further details of the models and methods, including how to use this DNN model to make predictions, are documented in SM §3.

**Conclusions**

We developed a data-driven approach that combined large-scale atomistic simulations, GA-based variable selection, and DNN to predict the temperature- and bulk composition-dependent GB $\Gamma_{\text{Ag}}$, $\Gamma_{\text{Disorder}}$, and $V_{\text{Free}}$ diagrams as functions of five GB crystallographic DOFs. These include not only the well-studied symmetric (ST and TW) GBs, but also the more general asymmetric (AT and MX) GBs that had been less studied before. The unified all-included DNN model in fact performs better for more general and asymmetric GBs.

To our best knowledge, this work has also generated the largest and most systematic (>6500 hybrid MC/MD) atomistic simulations dataset for binary GBs to date, thereby enabling the DNN model to predict the GB properties as functions of seven (five crystallographic plus two thermodynamic) DOFs. The GA-based variable selection discovered interesting characters for each of four GB groups, as well as for all GBs combined together. Finally, a unified all-included DNN model has been developed to predict the GB $\varGamma_{\text{Ag}}$, $\varGamma_{\text{Disorder}}$, and $V_{\text{Free}}$ diagrams, which are ~$10^8$ faster than the atomistic simulations.

**Acknowledgement**

This work is supported by a Vannevar Bush Faculty Fellowship sponsored by the Basic Research Office of the Assistant Secretary of Defense for Research and Engineering and funded through the Office of Naval Research under Grant No. N00014-16-2569. We thank Dr. Shengfeng Yang for the initial training and helpful discussion about using hybrid MC/MD simulations. Simulations and calculations were performed at the Triton Shared Computing Cluster (TSCC) at the University of California, San Diego (UCSD).

**Author contributions**

J. L. conceived the idea and supervised the work. C. H. performed simulations and calculations. C. H. and J. L. drafted the manuscript. S. P. O. supervised DNN development. Y. Z., C. C., and C. H. designed the architecture of DNN model. Y. Z. trained and tested the DNN models. All co-authors reviewed and revised the manuscript.

**Competing interests:** The authors declare no conflict of interests.

**Data availability:** Substantial data are documented in the 152-page long Supplementary Material.

**Table 1.** GA-selected significant GB descriptors based on three GB properties for four individual groups of GBs, as well as all-included GBs. $\Sigma$ is the GB coincidence number, and $\Sigma^{-1}$ is its reciprocal. $d_1$ and $d_2$ are the interplanar distances of Plane 1 and 2, and the effective interplanar distance for an asymmetric GB is defined as: $d_{\text{eff}} = (d_1 + d_1)/2$. $(r, \theta_{\text{sph}}, \varphi_{\text{sph}})$ is the the sphereical coodination of Miller index $(h\ k\ l)$. $(n\ m\ o)$ is the normal vector of Plane 1 or 2 (denoted as subscript). Tilt (decomposed tilt-rotation) or twist (decomposed twist-rotation) axis is represented by $[u_{\text{tlt}}\ v_{\text{tlt}}\ w_{\text{tlt}}]$ or $[u_{\text{twst}}\ v_{\text{twst}}\ w_{\text{twst}}]$, with the corresponding tilt or twist angle ($\varphi_{\text{tlt}}$ or $\Phi_{\text{twst}}$). The misorientation angle is $\theta_{\text{mis}}$. The overall (undecomposed) rotation axis is represented $[u_{\text{rot}}\ v_{\text{rot}}\ w_{\text{rot}}]$ with the rotation plane $(h_{\text{rot}}\ k_{\text{rot}}\ l_{\text{rot}})$. The empty space denotes that no significant GB descriptor was elected by the GA in this sub-group. See detailed descriptions in §1.6 and Table S2 in SM.

| GB Descriptors | Symmetric-tilt (ST) | Twist (TW) | Asymmetric-tilt (AT) | Mixed (MX) | All-included (All) |
|---|---|---|---|---|---|
| GB coincidence parameter | $\Sigma$ | $\Sigma$ | $\Sigma$ | | $\Sigma^{-1}$ |
| Interplanar distances | | | $d_1$ | | $d_1$ |
| Miller index of Plane 1 ($h_1\ k_1\ l_1$) | $h_1, l_1$ | | $h_1$ | $h_1$ | $k_1, l_2$ |
| Miller index of Plane 2 ($h_2\ k_2\ l_2$) | $l_2$ | | | $k_2$ | $l_2$ |
| Sphereical Plane 1 ($r_1, \varphi_{\text{sph}}^1, \theta_{sph}^1$) | | $r_1$ | | $\theta_{\text{sph}}^1$ | $\theta_{\text{sph}}^1, \varphi_{\text{sph}}^1$ |
| Sphereical Plane 2 ($r_2, \varphi_{\text{sph}}^2, \theta_{sph}^2$) | $\theta_{\text{sph}}^2$ | $r_2$ | | | $\varphi_{\text{sph}}^2$ |
| Normal of Plane 1 ($n_1\ m_1\ o_1$) | $n_1$ | | | | |
| Normal of Plane 2 ($n_2\ m_2\ o_2$) | $m_2, o_2$ | | $n_2$ | $n_2$ | $n_2, o_2$ |
| Tilt-rotation axis $[u_{\text{tlt}}\ v_{\text{tlt}}\ w_{\text{tlt}}]$ | $u_{\text{tlt}}, v_{\text{tlt}}, w_{\text{tlt}}$ | | $u_{\text{tlt}}, v_{\text{tlt}}, w_{\text{tlt}}$ | $u_{\text{tlt}}, w_{\text{tlt}}$ | |
| Twist-rotation axis $[u_{\text{twst}}\ v_{\text{twst}}\ w_{\text{twst}}]$ | | | | | |
| Angles: $\varphi_{\text{tlt}}, \Phi_{\text{twst}}, \theta_{\text{mis}}$ | $\varphi_{\text{tlt}}, \theta_{\text{mis}}$ | $\Phi_{\text{twst}}, \theta_{\text{mis}}$ | $\theta_{\text{mis}}$ | $\varphi_{\text{tlt}}$ | $\varphi_{\text{tlt}}$ |
| Rotation axis $[u_{\text{rot}}\ v_{\text{rot}}\ w_{\text{rot}}]^*$ | | | | | $u_{\text{rot}}, v_{\text{rot}}$ |
| Rotation plane $(h_{\text{rot}}\ k_{\text{rot}}\ l_{\text{rot}})^*$ | | | | $h_{\text{rot}}, k_{\text{rot}}, l_{\text{rot}}$ | $k_{\text{rot}}, l_{\text{rot}}$ |

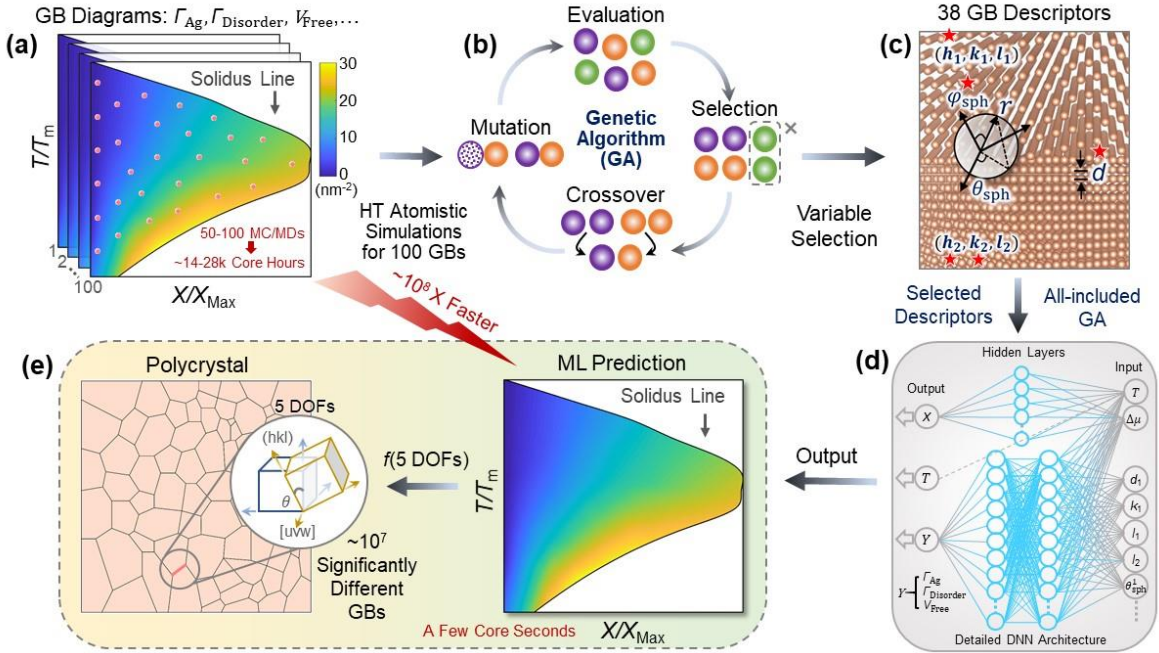

**Figure 1.** Workflow of machine learning prediction of bulk composition- and temperature-dependent grain boundary (GB) diagrams as a function of five macroscopic degrees of freedom (5 DOFs). **(a)** 6,581 individual isobaric semi-grand-canonical (constant-$N\Delta\mu PT$) ensemble atomistic simulations were performed for 100 representative GBs to calculate three types of GB diagrams of adsorption ($\Gamma_{Ag}$), excess disorder ($\Gamma_{Disorder}$), and free volume ($V_{Free}$). ~50-100 atomistic simulations (schematically represented by the red dots) are generally required to interpolate one set of three GB diagrams, which takes around 14,000-28,000 core hours of the simulation time per GB. **(b)** Schematic illustration of genetic algorithm (GA) based selection of significant GB descriptors. **(c)** The selected significant GB descriptors (indicated by red pentagram stars) were used as the input parameters to train, evaluate, and test deep neural network (DNN) models. **(d)** Schematic diagram of a two-layer single-task DNN with a 15-18-10-1 architecture for predicting GB properties ($Y = \Gamma_{Ag}$, $\Gamma_{Disorder}$, and $V_{Free}$) combined with a simplified single-layer artificial neural network (ANN) for predicting the bulk (grain) atomic fraction of Ag ($X_{Ag}$, which is normalized to the maximum solubility and represented by $X/X_{Max}$). The input parameters for the all-included DNN mode are the significant GB descriptors selected by GA plus two thermodynamic DOFs ($\Delta\mu$ and $T$). Here, the ANN predicted bulk (grain) composition at given $\Delta\mu$ and $T$ is a bulk property, independent of the GB structure. **(e)** The established DNN model can predict GB diagrams in a few core seconds per GB (*i.e.* ~$10^8$ faster than the atomistic simulation). This DNN-based machine learning model enables the forecast of the $\Gamma_{Ag}$, $\Gamma_{Disorder}$, and $V_{Free}$ diagrams of millions of distinctly different GBs as a function of five macroscopic DOFs, which is otherwise a "mission impossible" to construct by either experiments or modeling.

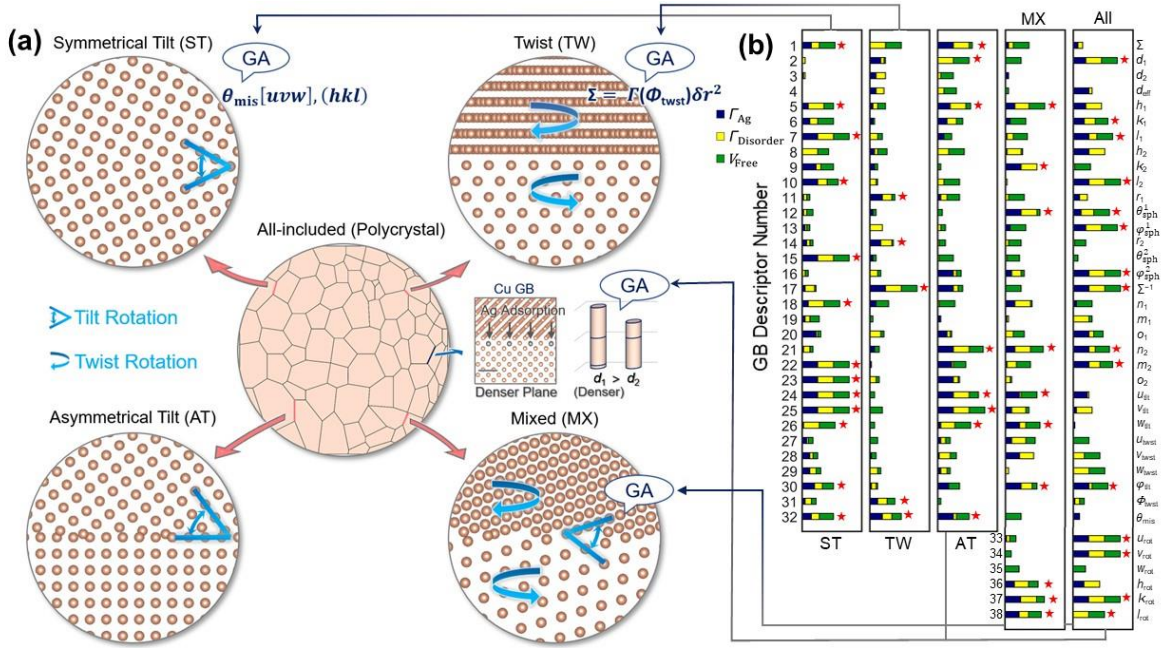

**Figure 2.** The classification of grain boundaries (GBs) and genetic algorithm (GA) based variable selection for GB descriptors. **(a)** GBs in a polycrystal can be classified into four types: symmetric-tilt (ST), twist (TW), asymmetric-tilt (AT), and mixed tilt-twist (MX) GBs. **(b)** Plots of the GA scores for 32 GB descriptors for ST, TW, and AT GBs, and 38 GB descriptors for MX and all four type GBs together (denoted as "all-included" or "All"). See the detailed descriptions for the 38 descriptors in Table S2 in SM. The red pentagram stars are used to label GA selected significant descriptors. The most significant GB descriptors selected by the GA include the parameters in the common notation $\theta_{\text{mis}}[uvw](hkl)$ for ST GBs and the those in the characteristic relation $\Sigma = \gamma(\Phi_{\text{twst}})\delta r^2$ for twist GBs. Moreover, the GA finds $d_1$ (of the lower-index plane) and $d_2$ to be the most and second most significant descriptors for AT GBs, as well as all-included GBs, which suggest that GB properties are dominated by the (denser) lower-index plane.

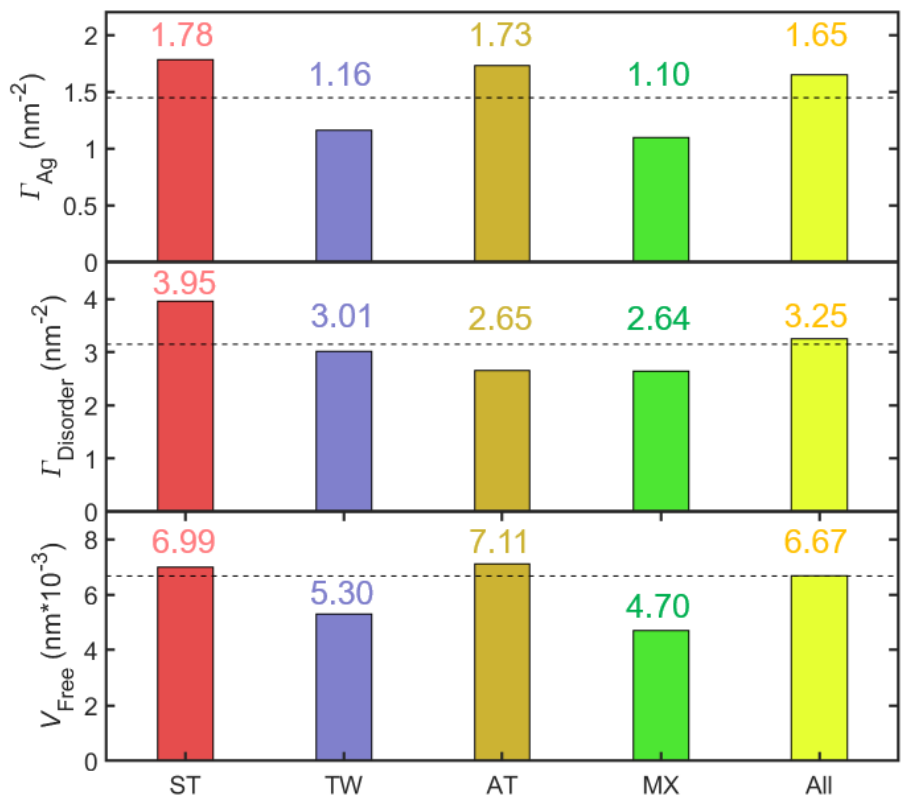

**Figure 3.** Comparison of weighted average root-mean-square errors (RMSEs) between the MC/MD simulations and DNN predictions of $\Gamma_{Ag}$, $\Gamma_{Disorder}$, and $V_{Free}$ for four individual models (ST, TW, AT, and MX) and the unified all-included (All) model. For each of the five DNN models, the weighted average RMSEs were calculated based on the RMSEs of training, validation, and test sets; the detailed histograms are shown in Fig. S11. The dashed lines represent the overall weighted average RMSEs of all four individual models combined.

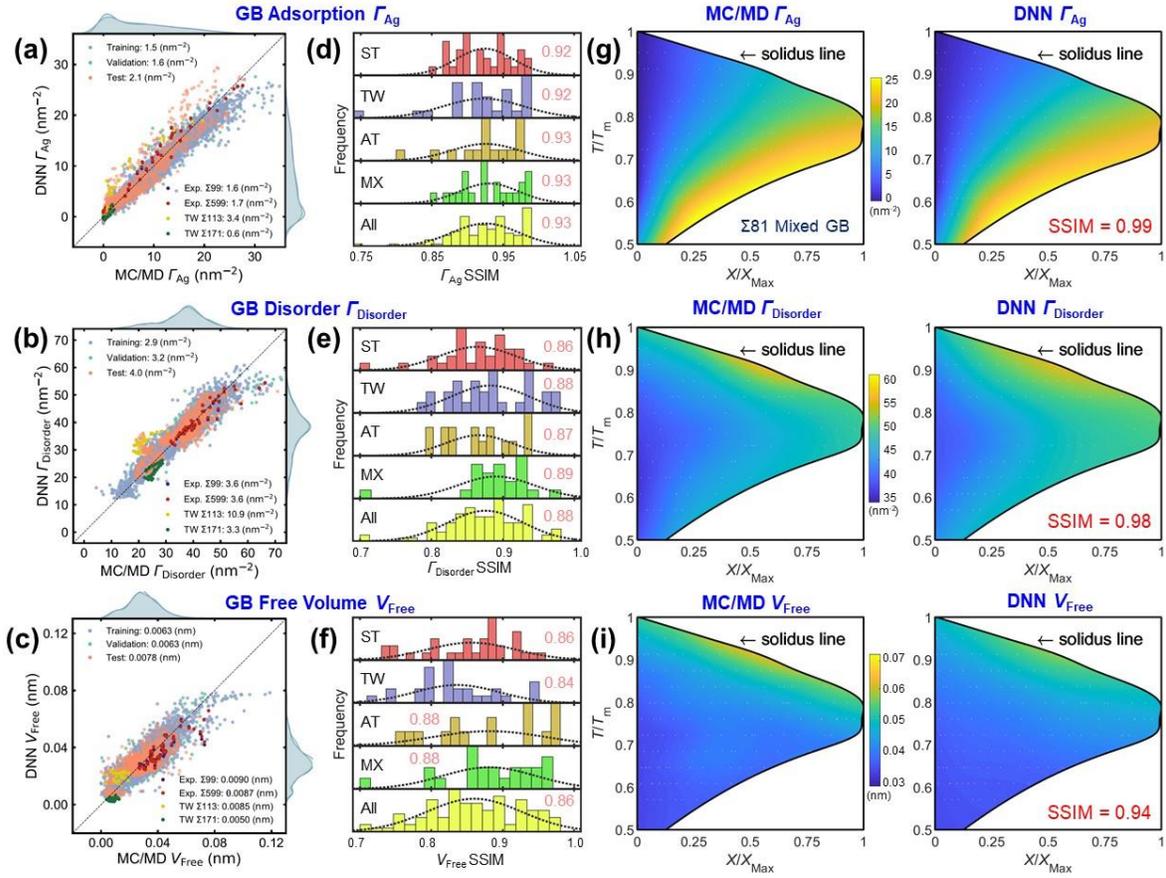

**Figure 4.** Performance of deep neural network (DNN) models. **(a-c)** Parity plots of DNN-predicted values of $\Gamma_{Ag}$, $\Gamma_{Disorder}$, and $V_{Free}$ using the all-included DNN model with an optimized network architecture 15-18-10-1 *vs.* hybrid MC/MD-simulated values (via $N\Delta\mu PT$ atomistic simulations). **(d-f)** Histogram plots with distribution line (black dotted line) of structural similarity index (SSIM) for characterizing the similarities between MC/MD-simulated GB diagrams and DNN-predicted GB diagrams of $\Gamma_{Ag}$, $\Gamma_{Disorder}$, and $V_{Free}$. The pink numbers labelled are mean SSIMs for each GB type. **(g-i)** Comparison of the MC/MD-simulated *vs.* DNN-predicted GB $\Gamma_{Ag}$, $\Gamma_{Disorder}$, and $V_{Free}$ diagrams using all-included DNN model for a Σ81 mixed GB with boundary planes $(1\bar{1}0)//(7\bar{8}7)$. The structural similarity indices (SSIMs) for characterizing the similarities between MC/MD-simulated and DNN-predicted GB diagrams are labeled.

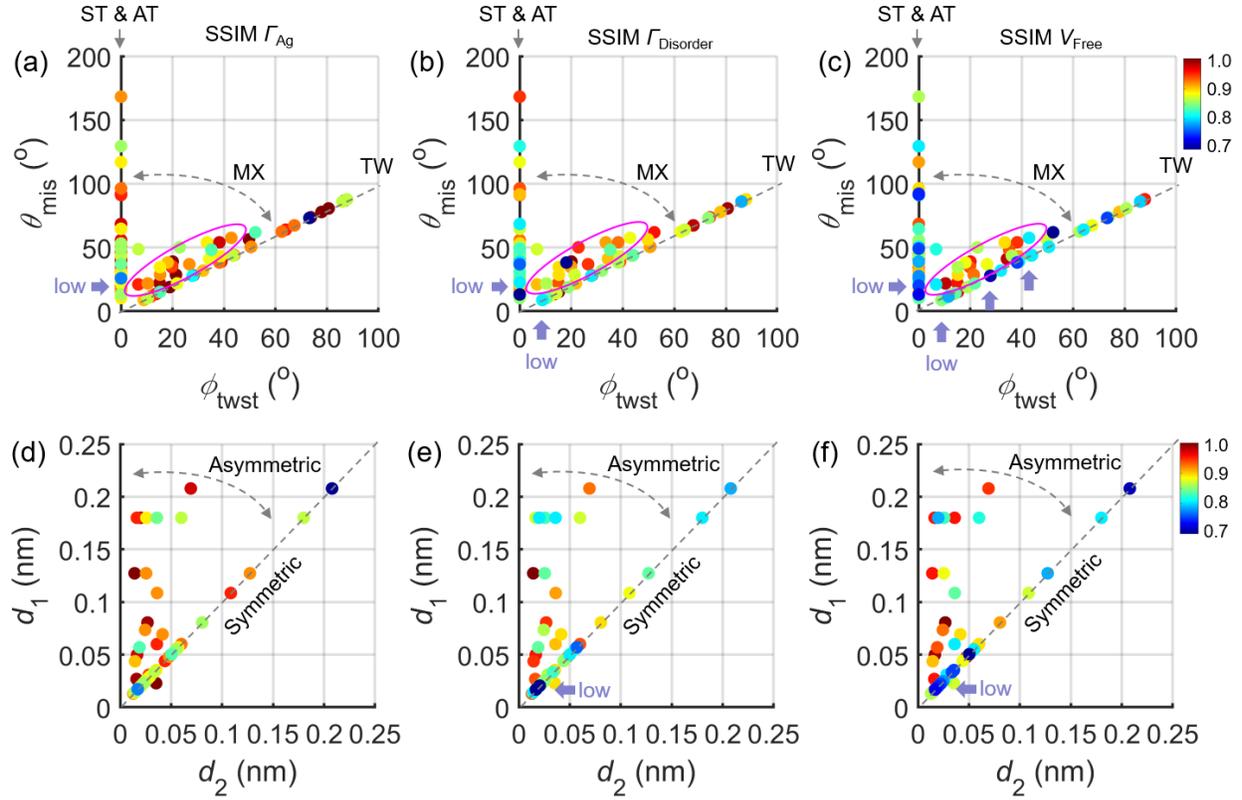

**Figure 5.** The SSIM distributions of 100 GBs for the **(a)** $\Gamma_{Ag}$, **(b)** $\Gamma_{Disorder}$, and **(c)** $V_{Free}$ diagrams plotted in the space of misorientation angle $\theta_{mis}$ and twist-rotation angle $\Phi_{twst}$. The dashed lines represent TW GBs ($\theta_{mis} = \Phi_{twst}$). The ST and AT GBs have $\Phi_{twst} = 0$, as indicated by the grey arrows. The MX GBs often have $\theta_{mis} \neq \Phi_{twst}$, which are indicated by dashed curves. The purple ellipses label the high SSIM values for MX GBs. The blue arrows indicate relatively low SSIM values for some tilt and twist GBs with low rotation angles. The SSIM distributions of 100 GBs for the **(a)** $\Gamma_{Ag}$, **(b)** $\Gamma_{Disorder}$, and **(c)** $V_{Free}$ diagrams plotted in the space of interplanar distances $d_1$ and $d_2$. The symmetric ST and TW GBs, where $d_1 = d_2$, lie on the dashed lines (albeit a few AT and MX GBs also have $d_1 = d_2$). Most (but not all) asymmetric AT and MX GBs have $d_1 \neq d_2$, which are indicated by the dashed curves. The blue arrows label the relatively low SSIM values. The color bars are used to represent for SSIM values from 0.7 (blue) to 1 (red).